\documentclass[runningheads,a4paper]{llncs}

\usepackage[caption=false]{subfig}
\usepackage{amssymb}
\usepackage{graphicx}
\usepackage{algorithmic}
\usepackage{algorithm}
\usepackage{float}
\usepackage{amsmath}
\newfloat{algorithm}{t}{lop}

\usepackage{url}
\urldef{\mailsa}\path|{alfred.hofmann, ursula.barth, ingrid.haas, frank.holzwarth,|
\urldef{\mailsb}\path|anna.kramer, leonie.kunz, christine.reiss, nicole.sator,|
\urldef{\mailsc}\path|erika.siebert-cole, peter.strasser, lncs}@springer.com|    
\newcommand{\keywords}[1]{\par\addvspace\baselineskip
\noindent\keywordname\enspace\ignorespaces#1}

\include{Qcircuit}

\begin{document}

\mainmatter  

\title{Cross-level Validation of\\Topological Quantum Circuits}


\author{Alexandru Paler$^*$\and Simon Devitt$^+$\and Kae Nemoto$^+$\and Ilia Polian$^*$}
\authorrunning{Cross-level Validation of Topological Quantum Circuits}

\institute{$^*$University of Passau, Innstr. 43, 94032, Passau, Germany\\
$^+$ National Institute of Informatics, 2-1-2 Hitotsubashi, Chiyoda-ku, Tokyo, Japan
}

%
%

\toctitle{Lecture Notes in Computer Science}
\tocauthor{Authors' Instructions}
\maketitle

\begin{abstract}
Quantum computing promises a new approach to solving difficult computational problems, and the quest of building a quantum computer has started. While the first attempts on construction were succesful, scalability has never been achieved, due to the inherent fragile nature of the quantum bits (qubits). From the multitude of approaches to achieve scalability topological quantum computing (TQC) is the most promising one, by being based on an flexible approach to error-correction and making use of the straightforward measurement-based computing technique. TQC circuits are defined within a large, uniform, 3-dimensional lattice of physical qubits produced by the hardware and the physical volume of this lattice directly relates to the resources required for computation. Circuit optimization may result in non-intuitive mismatches between circuit specification and implementation. In this paper we introduce the first method for cross-level validation of TQC circuits. The specification of the circuit is expressed based on the stabilizer formalism, and the stabilizer table is checked by mapping the topology on the physical qubit level, followed by quantum circuit simulation. Simulation results show that cross-level validation of error-corrected circuits is feasible.

\end{abstract}

\section{Introduction}

Building a large scale quantum computer has been the focus of a large international effort for the past two decades.  
The fundamental principles of quantum information have been well established \cite{NC00} and experimental 
technologies have demonstrated the basic building blocks of a quantum computer \cite{LJLNMO10}. A significant 
barrier to large scale devices is the inherent fragility of quantum-bits (qubits) and the difficulty to accurately control 
them.  The intrinsic error rates of quantum components necessitates complicated error correction protocols  to be 
integrated into architecture designs from the beginning, and it's these protocols that contribute to the majority of 
physical resources (both in terms of total number of physical qubits and total computational time) necessary for 
useful algorithms.  

Topological Quantum Computation (TQC) \cite{FG08,RHG07} has emerged as arguably the most promising error 
correction model to achieve large scale quantum information processing.  This model incorporates a powerful error correction code
and has been shown to be compatible with a large number of physical systems \cite{DFSG08,JMFMKLY10}.  While experimental 
technology is not yet of sufficient size to implement the full TQC model, there have been demonstrations 
of small scale systems and no fundamental issue prevents further expansion to a fully scalable quantum computer.  

\begin{figure}[t!]
\centering

\subfloat[Original and compressed TQC circuits]
{
\includegraphics[width=6cm]{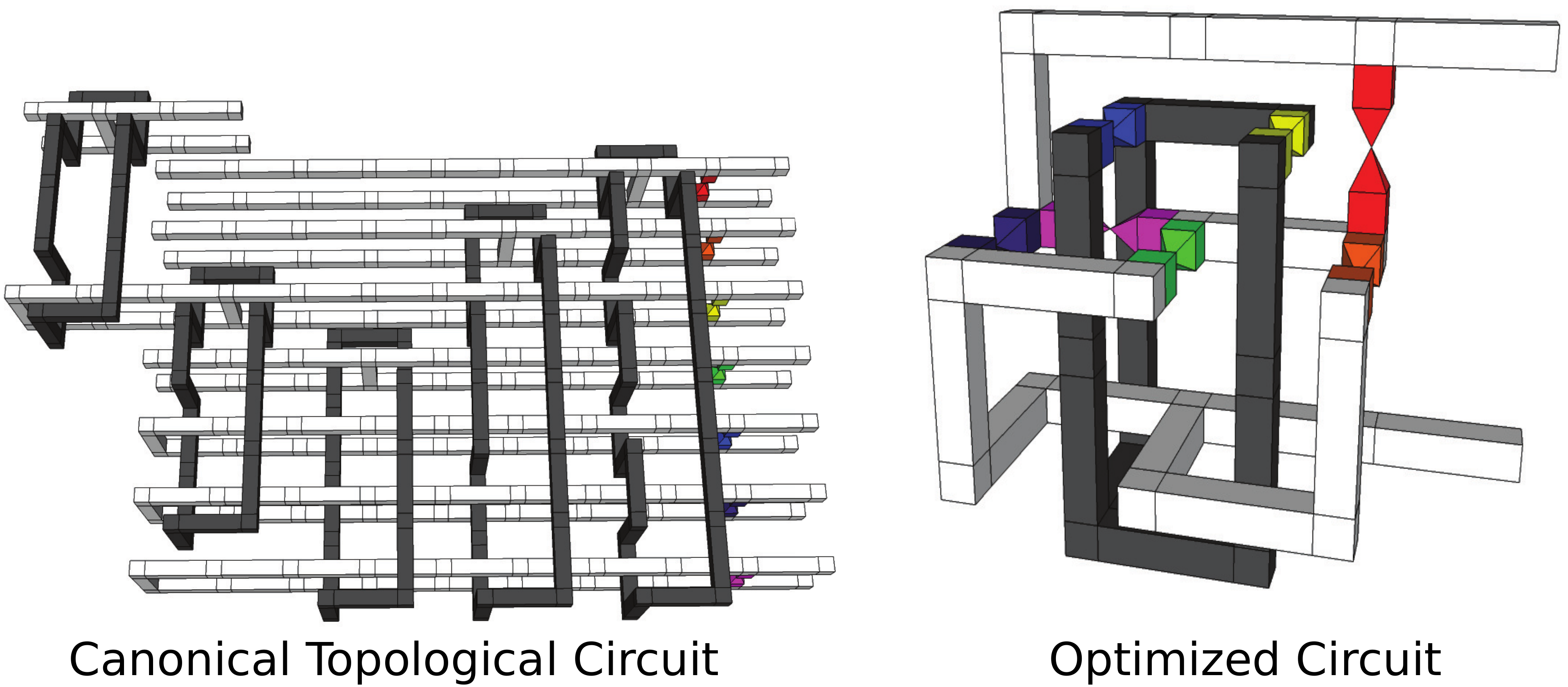}
\label{fig:pstate}
}%
\subfloat[The TQC design stack]
{
\includegraphics[width=4cm]{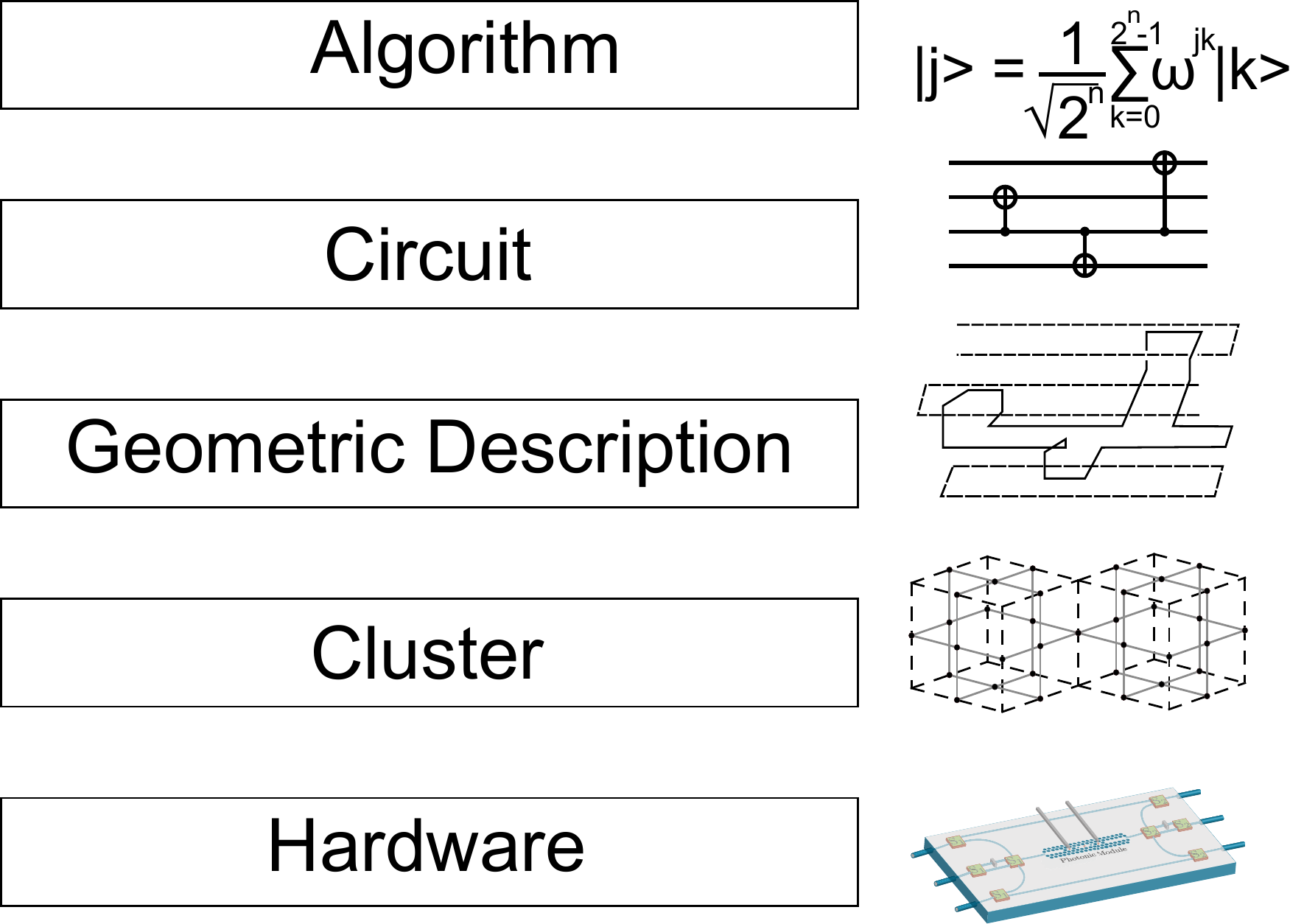}
\label{fig:levels}
}
\caption{Topological Quantum Computation (TQC)}
\end{figure}

The TQC hardware is responsible for producing a generic 3-dimensional lattice of qubits, 
and programming in the TQC model can be separated from the basic functionality of the quantum hardware.
Programming a TQC computer requires systematic methods, which are formulated starting from the 
TQC design stack (Figure \ref{fig:levels}) \cite{DFSG08}.
The stack consists of several abstraction levels that differ from the ones used in classical 
circuit design. The high level quantum algorithm is first decomposed into a quantum circuit. This circuit does not 
include any error correction protocols; these can be implemented in multiple ways, 
leading to circuits requiring a differing number of qubits and/or computational times.  
We then identify each qubit in the circuit, as logically encoded with the topological code. This transforms each \textit{logical 
qubit} into a large number of \textit{physical qubits} allowing for the implementation of correction protocols.  Such protocols 
also restrict the types of operations that can be performed on logical data, hence the quantum circuit needs to be 
further decomposed into gates from an universal set, but which can also be realized within the 
code.  Once these decompositions are complete, the resulting TQC circuit needs to be optimized 
with respect to the physical resources and then translated to the physical operations sent to the hardware.  

The qubit-lattice produced by the hardware embeds the topological quantum circuit and therefore it's physical size (volume) 
directly relates to the number of physical qubits employed for computation.  The computation can be 
constructed from the circuit in a straightforward, yet suboptimal, way \cite{FD12} (i.e. it will occupy a 3-dimensional volume much larger than 
required). The primary goal of TQC circuit synthesis is to construct an automated procedure that not only performs the 
required translation from circuit to topological circuit, but also to optimize the volume of these structures to ultimately 
reduce physical resources needed by the hardware. An example of an optimized circuit is presented in Figure \ref{fig:pstate}.

Validation of topological circuits is therefore a necessity, as optimized circuits often bare little resemblance to their 
original specification (e.g. Figure \ref{fig:pstate}). 
Validation has to be automated, as large topological circuits are complex objects, where the gate list is difficult to be extracted, and 
unfeasible to verify manually.

In this paper, we introduce the first automated validation method for TQC 
circuits. The input of the method is a quantum circuit specification, and the procedure verifies
that an instance of the quantum circuit exhibits the same functionality as the specification. 

For this purpose, we show that the validation problem
can be mapped to an equivalent problem that can be efficiently simulated. Direct simulation is necessary 
to confirm that the topological structure correctly implements the desired circuit.
Note that the simulator checks functionality of the topological structure, and it does not simulate 
error correction within the computation, as this is unnecessary for circuit validation.

\section{Background}

Quantum circuits are defined as series of quantum gates
applied to transform the state of qubits. Classical
bits can be either $0$ or $1$, while a qubit can have an infinity of states that can
be visually represented as points on the surface of a unit sphere (the
Bloch sphere). Quantum computing is based on
the postulates of quantum mechanics: the
state space of a quantum system (for our discussion a quantum
computer operating on $n$ qubits) is a complex space, where the system's
state is represented by unit vectors. For example, the state of a single
qubit is represented by a complex vector of length $2$, 
and the $2^n$-dimensional state of all $n$ qubits
is the tensor product of the component one-qubit states.
 The difficulty of
simulating a general quantum system using a classical computer stems 
from the exponential increase of the state representation
requirements. For example, the possible states of a two-qubit quantum
computer where each input qubit is initialized to the $\ket{0}=(1,0)^T$
state is represented using $2^2$ complex numbers $(1,0,0,0)^T$. The complex
entries of the state vector are called probability amplitudes, and arbitrary
tensor products of $\ket{0}$ and $\ket{1}=(0,1)^T$ (e.g. $\ket{000}$, $\ket{100}$,
$\ket{1111}$) are called computational-basis-states.

In the quantum circuit formalism, the evolution of the quantum computer's
state is dictated by the sequential application of quantum gates. The
state, after the application of each quantum gate, is modeled as the outcome of a
matrix-vector multiplication, thus the probability amplitudes of each
computational basis state are transformed. For this reason, quantum gates
can be understood as unitary complex matrices. Single-qubit quantum gates
are $2\times 2$ complex matrices, while $n$-qubit gates are $2^n \times 2^n$
complex matrices. The following gates are particularly relevant for our work.

\begin{eqnarray*}
\begin{array}{ccc}
X & = &
\left( \begin{array}{cc}
	0 & 1\\
	1 & 0
	\end{array} \right)
\\
 Z & = &
\left( \begin{array}{cc}
	1 & 0\\
	0 & -1
	\end{array} \right)
\end{array}
\,\,
\begin{array}{ccc}
H & = &
\frac1{\sqrt2}\left( \begin{array}{cc}
	1 & 1\\
	1 & -1
	\end{array} \right)
\\
T & = &
\left( \begin{array}{cc}
	1 & 0\\
	0 & e^{-i\pi/4}
	\end{array} \right)\\
\end{array}
\,\,
\begin{array}{ccc}
CNOT& = & 
\left( \begin{array}{cccc}
	1 & 0 & 0 & 0\\
	0 & 1 & 0 & 0\\
	0 & 0 & 0 & 1\\
	0 & 0 & 1 & 0
	\end{array} \right)
\end{array}
\end{eqnarray*}

A \emph{two-qubit controlled-gate} is applied to two
qubits, where one of the qubits is left unchanged, but controls (given its
state) the application of a single-qubit gate on to the second qubit. One
such gate is the $CNOT$ (Controlled-$X$) gate, where the first qubit 
is the control-qubit, and the second-qubit is the
target-qubit. Only when the control-qubit is
$\ket{1}$ the state of the target qubit is flipped (e.g. $\ket{0}$ becomes 
$\ket{1}$). Because of its action, the $X$-gate is called the bit-flip gate.

One of the major differences between classical and quantum computation 
is the concept of superposition. A qubit is a superposition, if more then one 
computational basis-state amplitudes is different than zero. The Hadamard gate can be
 used to construct the  $\ket{+}$ and $\ket{-}$ superpositions,
because $\ket{+}=H\ket{0} = \frac{1}{\sqrt{2}}(\ket{0}+\ket{1})$ and
$\ket{-}=H\ket{1} = \frac{1}{\sqrt{2}}(\ket{0}+\ket{1})$. Furthermore, the
state of at least two qubits is entangled if their composite state cannot be
written as a tensor product. For example, if the CNOT is applied to the
$\ket{0}\ket{+}=\ket{0+}$ state, the result
$\frac{1}{\sqrt{2}}(\ket{00}+\ket{11})$ is representing both a superposition
and an entangled pair of qubits. Similarly to the $X$-gate, the $Z$-gate is
called the phase-flip gate, because when applied to a single qubit it flips
the sign of the so-called relative phase (e.g. $\ket{+}$ is transformed into
$\ket{-}$).

In general, arbitrary quantum computations can be mapped to a
 discrete set of gates consisting of $\{(H,Z,X,T,CNOT\}$ with any desired accuracy. The $H$ gate is
used to construct superpositions, the $CNOT$ to construct entanglement and the
$T$ gate is used to achieve arbitrary single-qubit state rotations (visualized as
point rotations on the Bloch sphere surface).

\subsection{Stabilizer Formalism}

The exponential difficulty of describing the evolution of a quantum system
originates from the fact that, by incrementing the number of qubits operated on,
an exponential increase of the state-space is required. There is a particular
type of quantum computations for which this can be overcome by employing
the stabilizer formalism. Because $\ket{0}$ is an eigenvector with
eigenvalue $1$ of $Z$ it is said that $\ket{0}$ is stabilized by $Z$, and,
similarly, $\ket{1}$ is stabilized by $-Z$. Furthermore, using the same
idea, $\ket{+}$ is stabilized by $X$ and $\ket{-}$ is stabilized by $-X$.
Stabilizer circuits are circuits that can be decomposed into the gates
$\{X,Z,P,H,CNOT\}$ where $P=T\times T$. The identity matrix $I$ stabilizes any state,
while $-I$ is not a valid stabilizer. The state of such a circuit can be expressed
by its stabilizers, and it was shown that for $n$-qubit circuits $n$
stabilizers are required instead of $2^n$-dimensional complex amplitude vectors
\cite{NC00}. A stabilizer table $ST$ is an $n \times n$ table
consisting of $n$ independent stabilizers for the $n$ qubits of a computation (e.g. see
Figure \ref{fig:cnot}). The system's evolution of states is based on simple
transition rules (e.g. applying a $H$ gate on a qubit stabilized by $X$, results
in the state being stabilized by $Z$).

\begin{eqnarray*}
\textsc{Initial state:}\ket{+}\ket{+}\ket{0}&;  ST=\{XII, IXI, IIZ\}\\
\overset{H_1}{\rightarrow} \ket{0}\ket{+}\ket{0} &; ST=\{ZII, IXI, IIZ\}\\
\overset{CNOT_{2,3}}{\rightarrow} \ket{0}(\ket{00} + \ket{11})&; ST=\{ZII, IXX, IZZ\}\\
\end{eqnarray*}

The application of some gates, including the $T$ gate, cannot be expressed in a simple manner using
the stabilizer formalism. Its application to a state stabilized by $X$ results
in a state stabilized by a superposition
of stabilizers: $\frac{X + Y}{\sqrt{2}}$, where $Y=iXZ$. Thus, simulating a circuit with
$T$ gates using the stabilizer formalism requires doubling the set
of stabilizers each time a $T$ is encountered. The application of $T$ gates results in an exponential
increase of the state space to be observed. The set of stabilizing
gates together with the $T$ gate form an universal gate set, meaning
that an arbitrary quantum circuit can be expressed by its stabilizer
sub-circuits and a number of applications of $T$ gates (at the expense of an exponential
increase in computational resources).

\subsection{Measurement-based Quantum Computing}
\label{sec:22}

Arbitrary quantum computations can be mapped to the measurement-based quantum
computing paradigm (MBQC). MBQC utilizes an entangled ensemble of
qubits (\emph{cluster}) as a computational resource that is measured qubit-wise 
to perform quantum computations. During the measurement-process it is not 
 necessary to apply any entangling gates, because the cluster is used as the
entanglement resource.

In general, measuring a qubit is a probabilistic process
dictated by the probability amplitudes of its state. When a qubit is measured in
the computational basis (the $Z$-basis) the
qubit's state collapses to either $\ket{0}$ or $\ket{1}$, and when a qubit
is measured in the $X$-basis the possible outcomes are $\ket{+}$ and
$\ket{-}$. Furthermore, it is possible to perform \emph{rotated
measurements}, meaning that first the qubit's state is rotated and then an 
$X$- or $Z$-basis measurement is performed. In
measurement-based computing, the $T$ gate can be applied by using a 
rotated measurement. Two qubits 
$\ket{t}=\frac{1}{\sqrt{2}}(\ket{0}+r\ket{1})$ (where $r=e^{\frac{i\cdot\pi}{4}}$) and 
$\ket{q}=a\ket{0}+b\ket{1}$ are entangled using $CNOT$ resulting in
$\ket{tq}=a\ket{00}+ar\ket{11}+b\ket{01}+br\ket{10}$. 
The first qubit's $Z$-measurement will transform the
second qubit's state as if it were directly rotated by $T$: 
$a\ket{0}+r\ket{1}$ or $a\ket{1}+r\ket{0}$ (this result can be corrected using an $X$ gate) \cite{NC00}.

From the perspective of MBQC, only $X$- and $Z$-basis measurements
are necessary, iff the cluster to be measured contains already
rotated qubits (called injected qubits or \emph{injection points}). 
This is a technological detail that enables us to both simplify
the definition of the computing paradigm, and also to limit the number of
qubit states from the initial cluster to only two states:
$\ket{+}$ and $\ket{A}=\frac{1}{\sqrt{2}}(\ket{0}+e^{i\frac{\pi}{4}}\ket{1})$.

In the context of MBQC, the observation, that arbitrary circuits are formed by
stabilizer sub-circuits and applications of $T$ gates, can be further refined by
noting that arbitrary circuits are formed by only a stabilizer sub-circuit
(responsible for entangling the cluster-qubits) and another sub-circuit for
measuring the cluster-qubits.

\subsection{Topological Quantum Computation}

 One of the most promising approaches to construct 
a practical scalable fault-tolerant quantum computer, is based on the
topological error-correction code. This code lays at the foundation of
topological quantum computing (TQC), which is a measurement-based quantum
computing model. In the following a very short introduction to TQC
will be offered, while more details are to be found in \cite{RHG07,FG08}.

The TQC cluster has a repeating 3D graph
structure, which is obtained by stacking a \emph{unit-cell} along the three axis
(width, height and time). The temporal axis is dictated by the order of
performing the measurements. The unit-cell is constructed from $18$ physical
qubits (initialized into $\ket{+}$) and entangled using the Controlled-$Z$ gate
according to the pattern indicated in Figure \ref{fig:lclust}. Morever, for example,
by constructing a $2 \times 2 \times 2$ cluster of unit-cells, in the middle of the cluster
another unit-cell arises. The initial $8$ cells are known as \emph{primal cells},
and the central cell is called a \emph{dual cell}. 

Logical qubits are encoded into the cluster by disconnecting individual cluster-qubits
(achieved via $Z$-basis measurement).
Logical qubits are defined as pairs of \emph{defects}, 
where each defect is a \emph{trail of ''disconnected'' physical cluster-qubits},
and furthermore it can be geometrically abstracted (e.g. Figure \ref{fig:pstate}).
Cluster defects introduce degrees of freedom into the cluster, allowing for the storage
of error-protection information. Due to the duality of the
graph-structure, two types of logical qubits can be encoded: primal
and dual logical qubits, depending on whether qubits are removed from the primal or
the dual space. 


A logical qubit has a quantum state
which is protected against the errors, and the quantum gates can be implemented in 
a fault-tolerant manner directly on the logical qubits. The logical CNOT gate is always defined on
 logical qubits of opposite types, but it is still possible to define a logical CNOT between qubits
of the same type by using the circuit identities presented in \cite{FG08}.
Initializing and measuring logical qubits is performed by constructing 
the defect geometries presented in Figure \ref{fig:minit}.

\begin{figure}
\centering
\subfloat[]
{
\includegraphics[width=5cm]{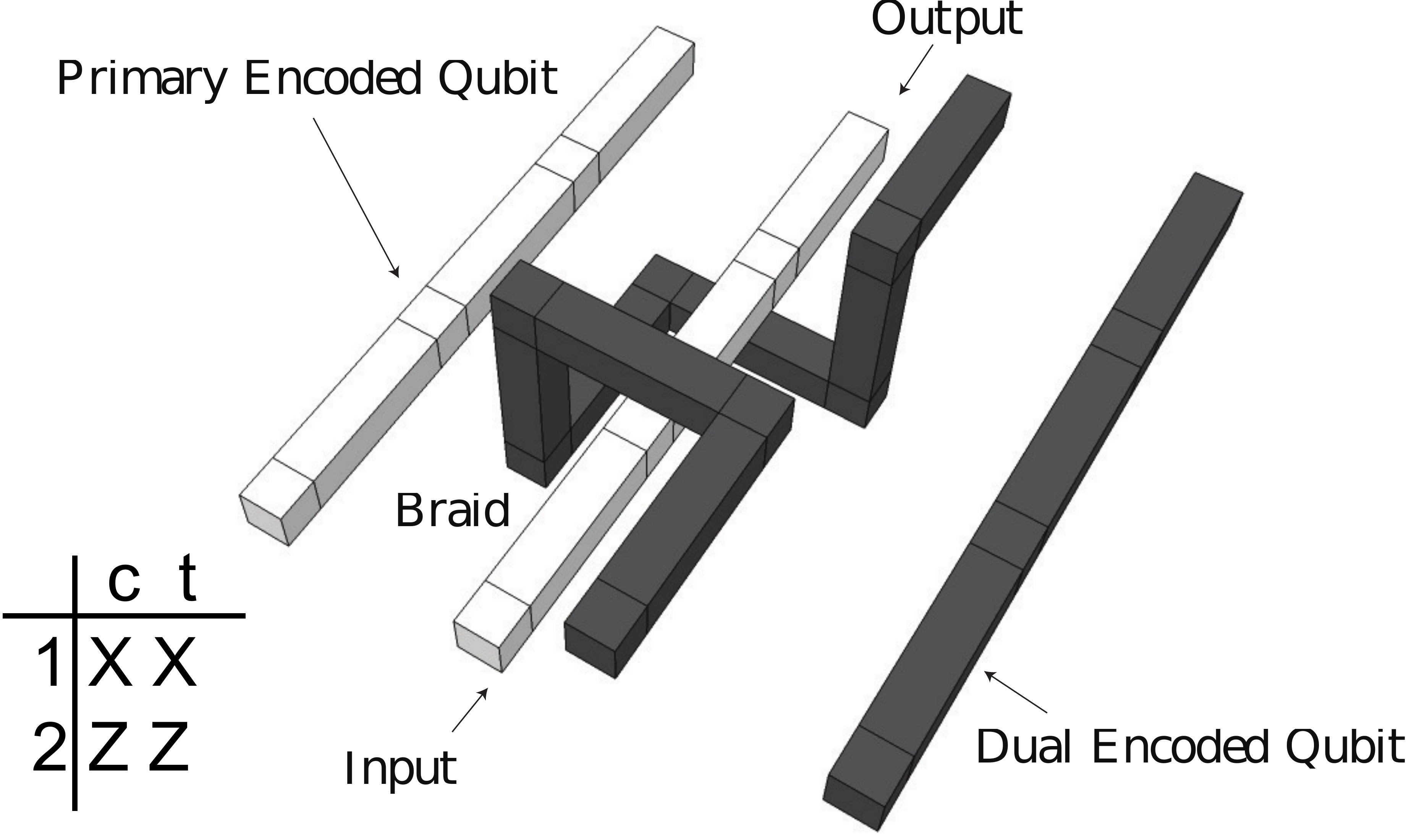}
\label{fig:cnot}
}%
\subfloat[]
{
\includegraphics[width=8cm]{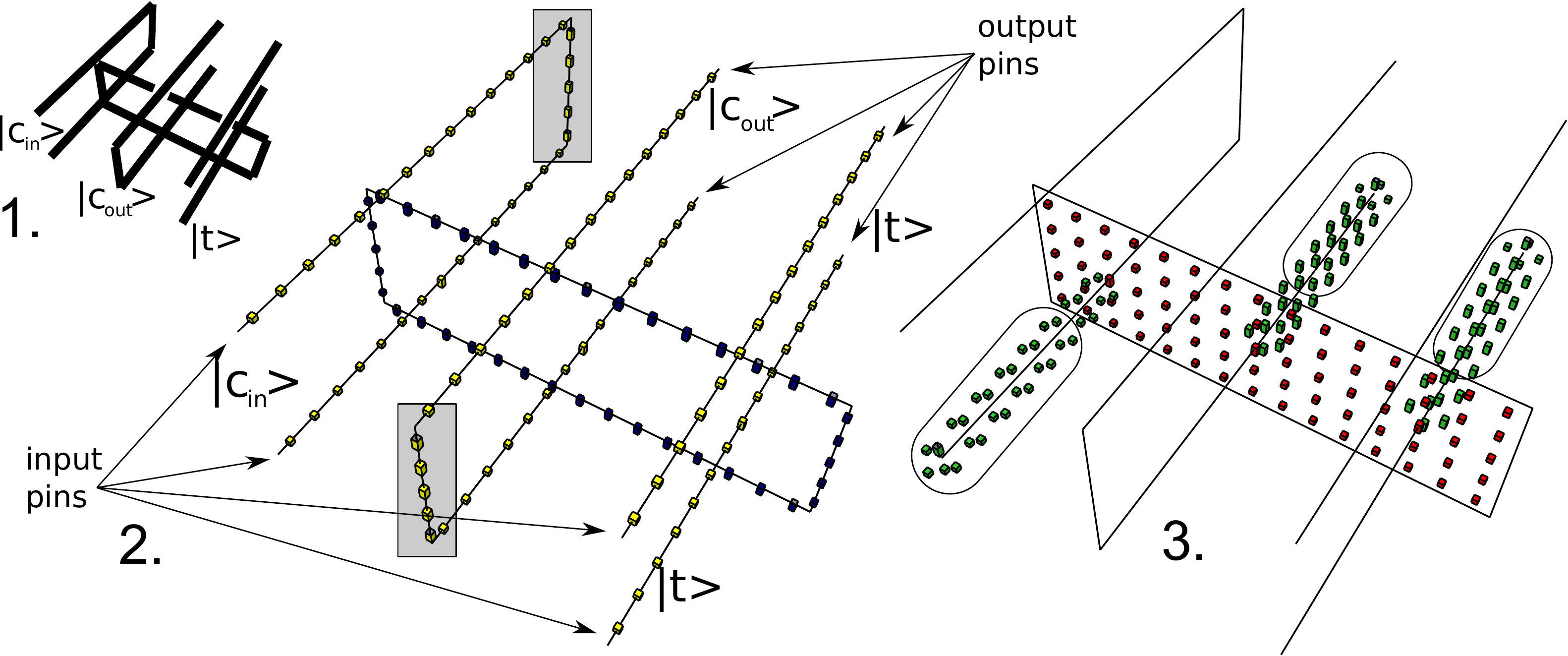}
\label{fig:example}
}
\caption{The logical CNOT: a) Two pairs of defects of opposite type are braided. The stabilizer table consists of two stabilizers and indicates, for example, that if the control-qubit is stabilized by $X$, after applying the CNOT the target-qubit will also be stabilized by $X$. b) Validation of a circuit consisting of 3 logical CNOTs: 1. the geometric description; 2. the mapped defect geometry, where $Z$-measured cluster qubits are indicated along with the input and output pins; 3. the correlations surface for the verification of one of the stabilizers from the specification.}
\end{figure}

A \emph{correlation surface} is a stabilizer defined over the cluster qubits that connect
the logical operators of the circuit's inputs to the logical operators of the
outputs, such that information is propagated correctly during the circuit 
operation \cite{FG08}. The
geometrical arrangements of the physical cluster qubits forming a
correlation surface are of two possible types: sheets and tubes (see Figure
\ref{fig:example}), and the physical cluster-qubits will be always measured in the $X$-basis. 
Sheets are spanned between logical qubit defects, while tubes encircle a given defect.
The cumulative parity of their measurement indicates how the logical stabilizers
of the logical qubits are to be interpreted. The measurement parity of a 
correlation surface is defined starting from the measurement results of the 
physical qubits in the surface. The measurement results
of an individual qubit are eigenvectors, with associated eigenvalues, of the 
measurement operator, and $1$ and $-1$ are the two possible eigenvalues for the $X$-measurement.
The \emph{measurement parity} along a correlation surface is the product
of the resulting associated eigenvalues. Finding a
correlation surface that connects the logical operators is not to be further
detailed into this work, because the methods enabling it are explained in
\cite{FG08}.

In TQC the computational universality is achieved by employing
injection points in a similar way how the rotational gate $T$ is applied by
teleportation as introduced in the context of MBQC. The TQC injection points are cluster qubits
initialized into the $\ket{A}$  state (defined in Section \ref{sec:22}) or $\ket{Y}=\frac{\ket{0} + i\ket{1}}{\sqrt{2}}$ state.
 Because TQC is an instance of MBQC, logical gate teleportation is achieved by
measuring the logical qubits that encode injected states.

\section{Validation of TQC Circuits}
\label{sec:val}



In order to formulate the cross-level validation of TQC circuits, we 
start with a consideration of generic (non-TQC) measurement-based fault-tolerant
quantum circuits. An arbitrary quantum circuit can be mapped to a 
construction from a stabilizer sub-circuit followed by a non-stabilizer sub-circuit 
that contains only rotated measurements. An adequate MBQC-oriented 
specification of such a ''decomposed'' quantum circuit is the tuple $QCS = \{ST, J, M\}$, where
$ST$ is the stabilizer table of the stabilizer sub-circuit, $J$ is the set
of injection points, and $M$ is the ordered set of measurements of these 
injection points. Given an implementation $QC$ that is also mapped to a tuple
$\{ST', J', M'\}$, we are interested in equivalence of both descriptions ($QC 
\equiv QCS$). If we assume that the number of injection points and their
measurement is not changed, as it will directly affect the computation 
being performed, this question is reduced to the equivalence checking of the
stabilizer circuit parts ($ST \equiv ST'$), which has previosusly been
investigated in the context of reversible computing \cite{wille2009equivalence}.

However, checking the equivalence of a TQC description against the 
specification $QCS$ is more challenging because no complete procedure to translate
the geometric description of the topological circuit to the stabilizer table
is currently known. In the 
following, we outline the cross-level approach which checks equivalence without 
constructing the stabilizer table.

\subsection{Problem Statement}

In the context of TQC, the stabilizers and the gates are defined at a 
logical level, which is constructed on top of the cluster-state level (physical 
qubits). The specification of the circuit ($QCS$ or, more exactly, the stabilizer 
table of its portion $ST$) refers to the logical level. In order to check the 
equivalence of the geometric description against $QCS$, we map the logical qubits to the
cluster state and validate it by simulation. This is 
done in two steps. First, the geometrical description is mapped to an 
(unmeasured) cluster. The mapping method can be derived from \cite{paler2014mapping}, and the details
are omitted here. Then, for every entry of the stabilizer table $ST$ from the
specification, the topological computation in the cluster is simulated 
using a (stabilizer) quantum circuit simulator. Note that the simulated geometry
is largely given by the shapes of the logical qubits which are independent from the
processed $ST$ entry. Moreover, the $ST$ entry determines the initialization and
measurement parts of the logical qubits (see Figure \ref{fig:minit}).

In the following paragraphs the validation procedure will be detailed and
analyzed.

\subsection{Validation Procedure}

The cross-validation of circuits is a simulation based procedure of a cluster
where the geometric description of the TQC circuit was mapped. 
Algorithm \ref{alg:1} is synthesizing the details that are presented in the following.

\begin{figure}[t!]
\centering
\subfloat[]
{
\includegraphics[width=6cm]{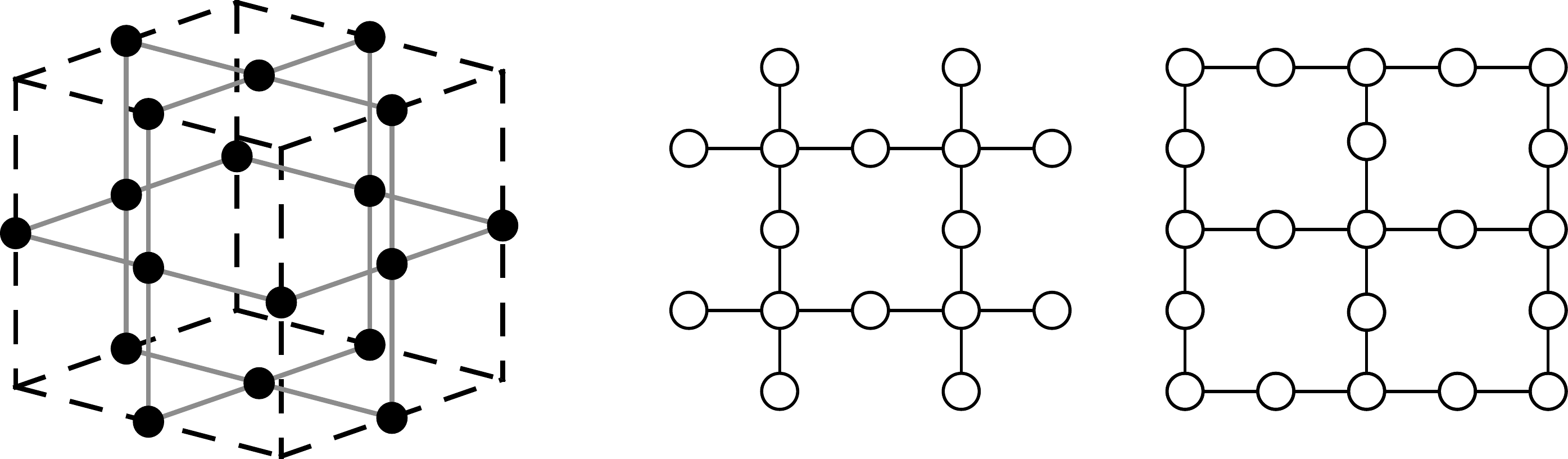}
\label{fig:lclust}
}%
\hfil
\subfloat[]
{
\includegraphics[width=4cm]{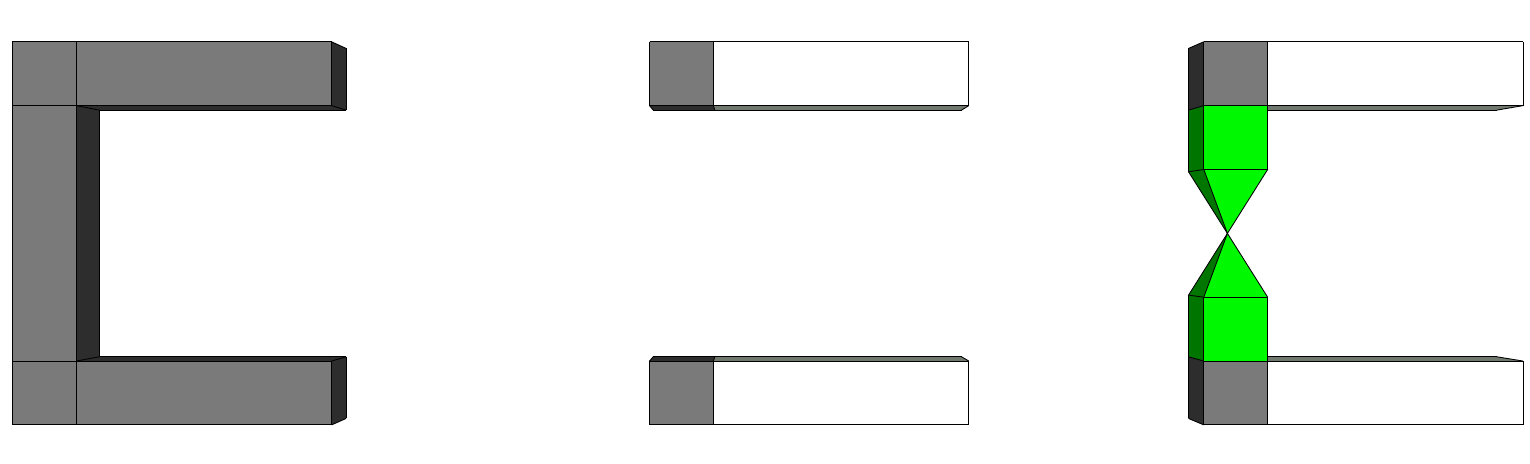}
\label{fig:minit}
}
\caption{TQC constructs: a) the unit-cell of 18-entangled qubits, and the two repeating layers that are simulate. b) Defect geometries for initialization of primal logical qubits: 1. $Z$-basis initialisation 2. $X$-basis initialisation 3. injection point. The defect geometries for measurement are similar.}
\end{figure}

The validation method starts by mapping the geometry to a cluster (Lines \ref{src:1}, \ref{src:2}). The set $TQCC=\{(x,y,z) |\;x,y,z \in \mathbb{N}, measure(x,y,z) \in \{X,Z\},$ $init(x,y,z) \in \{\ket{+}, \ket{A}, \ket{Y}\}\}$ is specified as a finite set of associated coordinates of physical qubits, that are marked for measurement in the $X$- or $Z$-basis and initialisation into $\ket{+}$, $\ket{A}$ or $\ket{Y}$. The 3D-coordinates correspond to the geometry presented in Figure \ref{fig:lclust}. Mapping of defect geometries to the 3D lattice takes an initial cluster $TQCC$, where no measurements were marked, and updates it: $Z$-basis measurements for defect-internal physical qubits, and $X$-basis measurements for all others. Injection points (physical qubits initialized into $\ket{A}$ or $\ket{Y}$) are measured in the $X$-basis.

Logical qubits can contain injection points anywhere along the
geometric structure, and the target of validation method is to check that before the injection point will be measured, the logical
qubit is correctly stabilized. Otherwise the result of the rotated measurement will be faulty, and the whole
quantum computation is compromised. During the validation, as indicated in Section \ref{sec:val}, injection points do not need to be explicitly considered. Without affecting the correctness of the method, these are initialized into the $\ket{+}$ state. By \emph{re-interpreting} the injection points, the $TQCC$ set is transformed into
$TQCC^+=\{(x,y,z) | x,y,z \in \mathbb{N}, measure(x,y,z) \in \{X,Z,I\}, init(x,y,z) \in \{\ket{+}\}\}$ (Line \ref{src:2}).

Similar to classical circuits where input and output pins are used for the inputs and the outputs of the circuit, $P_{in} \subset TQCC$ is a set of cluster coordinates of the physical qubits used for initiliazing the logical qubits. The same applies for physical qubits used for logical measurement. These are used to read the information from the TQC circuit; their coordinates are contained in the set $P_{out} \subset TQCC$ representing the \emph{output pins}. The physical qubits from both sets are marked for either $X$- or $Z$-basis measurement, in order to respect the defect geometries from Figure \ref{fig:minit} (Line \ref{src:3}). Cluster injection points are elements of $P_{out}$. Circuit simulation will be performed for each for each logical stabilizer specified in $ST$, and the $P_{in}$ and $P_{out}$ sets will be constructed accordingly.

A mapped cluster is supporting a logical stabilizer if the correlation surface that connects the
corresponding input and output pins has even parity (Lines \ref{src:13}, \ref{src:14}). In the absence of errors (which is assumed 
during the validation of circuit functionality) the topology of the 3D-cluster guarantees that the measurement 
parity of all the unit-cell
face-qubits is even \cite{RHG07}. The existence of the logical stabilizer support is proven 
by computing the parity of a correlation surface, as even parity 
indicates that the stabilizer can be correctly constructed using physical-cluster qubits.

In order to check the existence of all the logical stabilizers specified in $ST$, the 
validation method checks each entry in the table sequentially (Lines \ref{src:4} -- \ref{src:14}).
 Depending on the logical stabilizer to be checked, 
the injection point coordinate will 
be marked in $TQCC^+$ (Line \ref{src:8}) with either an $X$-basis measurement
(if the logical qubit should be stabilized by logical-$X$) or with a $Z$-measurement (if the logical qubit should be stabilized by logical-$Z$).

Checking the support of a logical stabilizer is performed by simulations of the cluster,
and the simulation involves the following steps. The first step is to compute
the presumed correlation surface of the investigated stabilizer (Line \ref{src:7}). A correlation surface
connects, only for the logical qubits referenced by the stabilizer, the input to the output pins.
 In the second step all the physical qubits are measured according to their markings from $TQCC^+$.
In a third step, the existence of the stabilizer is determined based on the parity of the correlation surface
(Lines \ref{src:9} -- \ref{src:12}). The error-correction is neglected, as it does
not manifest itself in the validation of the specification $QC=\{ST, J, M\}$.

During the second step, the measurements are performed. This can be 
done in an arbitrary order, but we adopt a layered approach. 
One of the three dimensions of the cluster is defined to be the temporal axis. In a cluster of size $m \times n \times t$, 
we can reduce memory requirements by instead simulating a 
physical lattice of $t-1$ $m \times n \times 2$ \emph{layer pairs} of the cluster dynamically.

Each layer pair (Line \ref{src:15}) consists of two cross-sections of the cluster (e.g. Figure \ref{fig:lclust}).
Layer $i$
contains all physical qubits with $t$-coordinate equal to $i$. 

In the $i$-th simulation run ($i = 0, \ldots, t-1)$, layers $i$ and $i+1$ are considered.
The first simulation run considers only qubits from layers 1 and 2 with all
connections between these layers. However, $X$ and $Z$ measurements are
only performed on qubits with the $t$-coordinate 1.
In the second simulation run, the qubits from the second layer, which
retain their states from the first simulation run, are entangled with
(hitherto unconsidered) qubits from layer 3 (initialized to $\ket+$)
according to the unit-cell structure that is used throughout the complete
$m \times n \times t$ cluster. Only second-layer qubits are measured, which
influences the entangled third-layer qubits. This process is continued until the qubits of the
final layer $t$ are measured.

\begin{algorithm}
	\caption{Cross-level Validation}
\begin{algorithmic}[1]
	\REQUIRE{Circuit $TQC$ as a geometrical description and the specification $QCS$}
	\STATE{Compute $TQCC$ starting from the geometry of $TQC$} \label{src:1}
	\STATE{Compute $TQCC^+$ from $TQCC$ by marking injection points as $\ket{+}$ initialized} \label{src:2}
	\STATE{Compute $P^{q}_{in},P^{q}_{out} \subset TQCC^+$} \label{src:3}
	\FORALL{stabilizer $s$ from $ST$ of $QCS$} \label{src:4}
			\STATE{$SIMTQC \leftarrow TQCC^+$} \label{src:5}
			\FORALL{Logical qubit $q$ stabilized by $s$} \label{src:6}
				\STATE{Mark in $SIMTQC$ at $coord \in P^{q}_{in},P^{q}_{out}$ the geometric patterns for initialisation and measurement of $q$ according to $s$} \label{src:8}
				\STATE{Compute for $s$the correlation surface $CORS$} \label{src:7}
			\ENDFOR
			\STATE{$parity \leftarrow 1$} \label{src:9}
			\STATE{Construct layer $l_0$ of $SIMTQC$} \label{src:10}
			\FORALL{Layer $l_i$ of $SIMTQC$, $i>0$} \label{src:11}
				\STATE{Construct $l_{i}$ and Entangle with $l_{i-1}$}\label{src:15}
				\FORALL{Cluster qubits $cq$ in $l_{i-1}$, $cq \in CORS$}
					\STATE{$ev \leftarrow$ measure $cq$ in $X$-basis}
					\STATE{$parity = parity \cdot ev$}
				\ENDFOR
			\ENDFOR \label{src:12}
			\IF{$parity = -1$} \label{src:13}
				\RETURN{$TQC$ is NOT valid according to $QCS$}
			\ENDIF
	\ENDFOR
	\RETURN{$TQC$ is valid according to $QCS$} \label{src:14}
\end{algorithmic}
	\label{alg:1}
\end{algorithm}

\section{Results}
\label{sec:results}

To evaluate the practicality and the scalability of the validation procedure the quantum circuit simulator CHP \cite{AG04} was integrated for the cluster simulation step. Checking the complete stabilizer truth table $ST$ requires between $1$ and $|ST|$ simulations.

\begin{figure}
\centering
%
\includegraphics[width=8cm]{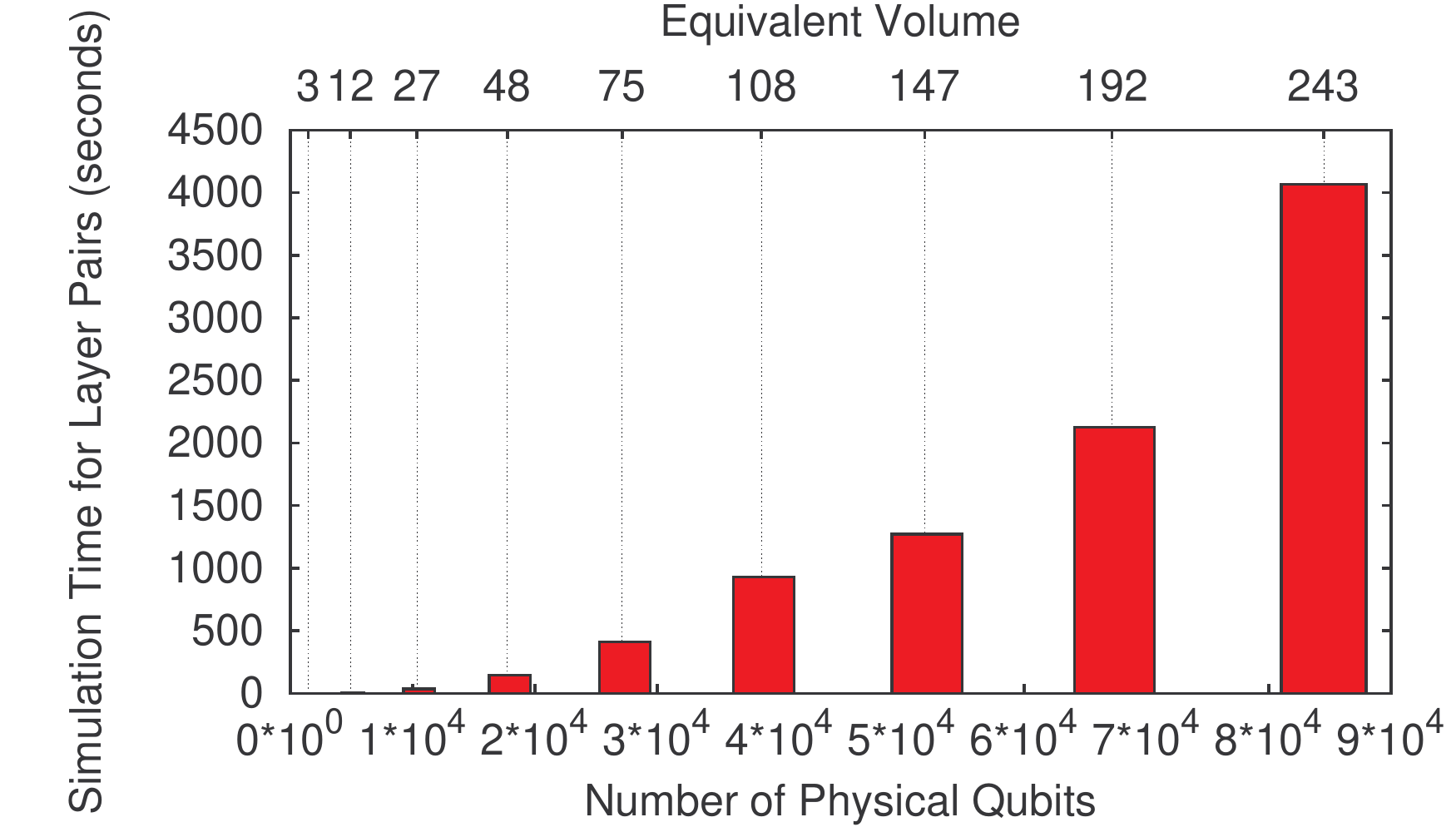}
\label{fig:sim1}
\caption{Average simulation times for pairs of layers}
\end{figure}

We considered TQC circuits consisting of logical $\textsc{CNOT}$ gates acting on logical qubits. Their sizes are expressed as an \emph{equivalent volume} \cite{FD12}, a quantity that measures the volume of a topological structure compared to a set of independent regularly stacked logical CNOT gates. Our results indicate that reduced TQC circuits of those equivalent volumes are feasible to simulate, and thus to validate. Average simulation times for one pair of layers in such circuits are reported in Figure \ref{fig:sim1}. For example, the number of physical qubits required to be simultaneously simulated for the circuit having the equivalent volume of three $\textsc{CNOT}$ gates was $1462$, and this number was $84,052$ for the equivalent volume of $243$ $\textsc{CNOT}$ gates. These results suggest that even large and complex topological quantum circuits can be validated in reasonable time.

\begin{figure}
\centering
\includegraphics[width=8cm]{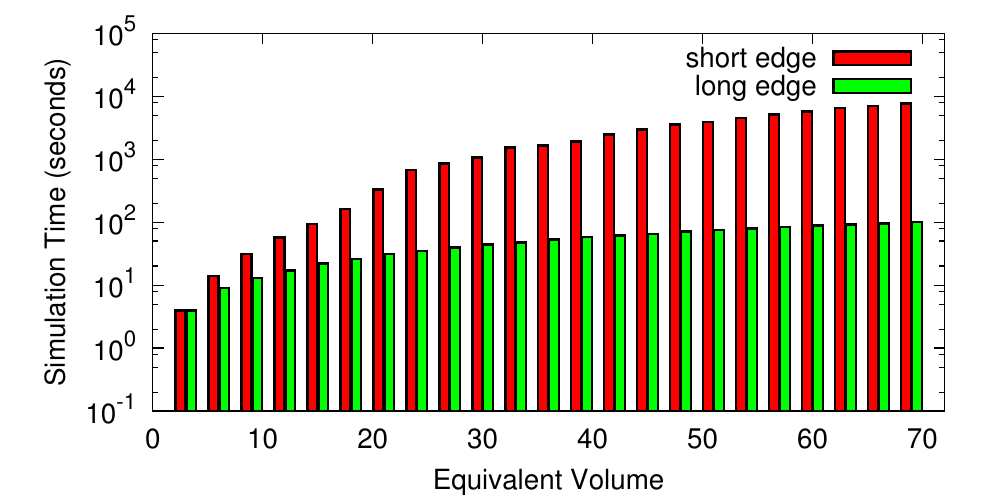}
\label{fig:sim2}
\caption{Simulation times after choosing different temporal axes}

\end{figure}

The selection of one of the three axes in the cluster as the temporal axis is arbitrary, which provides an additional degree of freedom for validation. The complete computation is confined to a 3D volume where the three edges may have different lengths. Selecting a short edge as the temporal axis will result in relatively small number of relatively large simulation instances, while selecting a long edge will require more simulations with less qubits per simulation. Note that the simulated functionality is identical for both options. Figure \ref{fig:sim2} compares the run times for these possibilities. It can be seen that simulation is orders of magnitude faster when the longest edge is selected. This is not surprising as the measurement of stabilizers is of quadratic complexity in the number of qubits, and therefore having to consider less qubits per simulation instance outweighs the higher number of simulation runs.

\section{Conclusion}

The first validation method for topological
quantum circuits was presented. Synthesis of
topological quantum circuits often results in non-obvious
inaccuracies that currently require a huge manual effort to find
and correct, which is clearly impractical
even for small circuits. The presented validation procedure maps
the geometric description to the actual three-dimensional cluster
of physical qubits and simulates these qubits. This abstraction
level is much closer to the actual hardware implementation and
is well suited to identify any deviations from the specification.
Empirical data show the scalability of the procedure to circuits
of practical size. As the next step, we plan to develop a
validation-guided synthesis procedure for topological quantum
circuits, and a more efficient representation of the circuit specification. 

\small
\bibliographystyle{splncs}
\bibliography{bib1.bib}

\end{document}